# A clock network for geodesy and fundamental science

C. Lisdat[1], G. Grosche[1]*, N. Quintin[2], C. Shi[3], S.M.F. Raupach[1], C. Grebing[1],
D. Nicolodi[3], F. Stefani[2,3], A. Al-Masoudi[1], S. Dörscher[1], S. Häfner[1], J.-L. Robyr[3],
N. Chiodo[2], S. Bilicki[3], E. Bookjans[3], A. Koczwara[1], S. Koke[1], A. Kuhl[1], F. Wiotte[2],
F. Meynadier[3], E. Camisard[4], M. Abgrall[3], M. Lours[3], T. Legero[1], H. Schnatz[1],
U. Sterr[1], H. Denker[5], C. Chardonnet[2], Y. Le Coq[3], G. Santarelli[6], A. Amy-Klein[2],
R. Le Targat[3], J. Lodewyck[3], O. Lopez[2], P.-E. Pottie[3]*

[1] Physikalisch-Technische Bundesanstalt, Bundesallee 100, 38116 Braunschweig, Germany.

[2] Laboratoire de Physique des Lasers, Université Paris 13, Sorbonne Paris Cité, CNRS,
99 Avenue Jean-Baptiste Clément, 93430 Villetaneuse, France.

[3] LNE-SYRTE, Observatoire de Paris, PSL Research University, CNRS, Sorbonne Universités,
UPMC Univ. Paris 06, 61 Avenue de l'Observatoire, 75014 Paris, France.

[4] Réseau National de télécommunications pour la Technologie, l'Enseignement et la Recherche,
23 – 25 rue Daviel, 75013 Paris, France.

[5] Institut für Erdmessung, Leibniz Universität Hannover, Schneiderberg 50, 30167 Hannover,
Germany.

[6] Laboratoire Photonique, Numérique et Nanosciences, UMR 5298 Université de Bordeaux 1,
Institut d'Optique Graduate School and CNRS, 1, Rue F. Mitterrand, 33400 Talence, France.

*Correspondence to: G.G. (email: gesine.grosche@ptb.de) and P.E.P. (email: paul-eric.pottie@obspm.fr).

Leveraging the unrivaled performance of optical clocks in applications in fundamental physics beyond the standard model, in geo-sciences, and in astronomy requires comparing the frequency of distant optical clocks truthfully. Meeting this requirement, we report on the first comparison and agreement of fully independent optical clocks separated by 700 km being only limited by the uncertainties of the clocks themselves. This is achieved by a phase-coherent optical frequency transfer via a 1415 km long telecom fiber link that enables substantially better precision than classical means of frequency transfer. The fractional precision in comparing the optical clocks of three parts in $10^{17}$ was reached after only 1000 s averaging time, which is already 10 times better and more than four orders of magnitude faster than with any other existing frequency transfer method. The capability of performing high resolution international clock comparisons paves the way for a redefinition of the unit of time and an all-optical dissemination of the SI-second.

Time and frequency are the most precisely measured physical quantities thanks to atomic clocks. Optical clocks are – due to their orders of magnitude better precision (*1*) and accuracy (*2,3,4*) than their microwave counterparts realizing the definition of the SI second – new sensors with highest resolving power to, e.g., test the temporal stability of fundamental constants (*5,6,7*). Complemented by phase-coherent telecom fiber links (*8,9,10,11*) as extremely powerful tools for frequency dissemination on a continental scale, new scenarios for fundamental and applied science will become realizable through large-scale fiber-based optical clock networks. Prominent examples are the search for dark matter (*12*) by monitoring the local time scale jitter when the Earth moves through cosmic domains. Also, accurately mapping the relative gravitational shift of clocks will enhance redshift tests (*13*) performed with clocks in space (ACES). Highly accurate clock networks can also perform tests of the Lorentz invariance (*14*) by searching for frequency modulation between clocks in a network moving in the Sun's gravitational potential, or can provide a better synchronization between very long baseline interferometers. Additionally, new geodetic reference frames based on relativistic geodesy (*15,16*) or a quantum network of clocks (*17*) can be established.

The traditional means to compare remote optical clocks are microwave ones: the optical clocks' frequencies are measured against a primary cesium microwave clock and the frequency values measured locally are compared, or satellite-based methods to transfer the frequency are applied (*18*). The accuracy limit for the former method is given by the cesium microwave clocks with the most stringent limit set by our groups at an agreement within $5 \times 10^{-16}$ (*7,19*); satellite link techniques do not reach significantly better performance either. Hence, the capabilities of distant optical clocks could so far neither be fully tested nor exploited. Here, we overcome this limitation by a direct, all-optical frequency comparison between two optical clocks via a telecom fiber link, demonstrate the first remote optical clock comparison with one order of magnitude better accuracy and orders of magnitude shorter averaging time than with microwave clocks or satellite links, and thus enable the above mentioned applications.

Furthermore, regular and practical international microwave clock comparisons are of utmost importance for the international time scale TAI (temps atomique international), which delivers the accuracy of primary frequency standards to a wide range of users in science and society. We show here for the first time that this scenario can be transferred to optical clocks and fiber links. This clears the path towards a redefinition of the unit of time, the SI second (*20,21*).

Our experimental setup is composed of two optical clocks and an optical frequency transfer system to allow their direct optical comparison. The two strontium optical lattice clocks $Sr_{SYRTE}$ and $Sr_{PTB}$ are located at the national metrology institutes LNE-SYRTE in Paris (*7*), France, and PTB in Braunschweig (*19*), Germany. At each institute, femtosecond frequency combs are used to accurately measure the frequency ratio of the optical clock to a transfer laser. Two stabilized fiber links transfer, respectively, the frequency information of the lattice clocks from Paris (*22*) and Braunschweig (*23*) to the connection point in Strasbourg (Fig. 1).

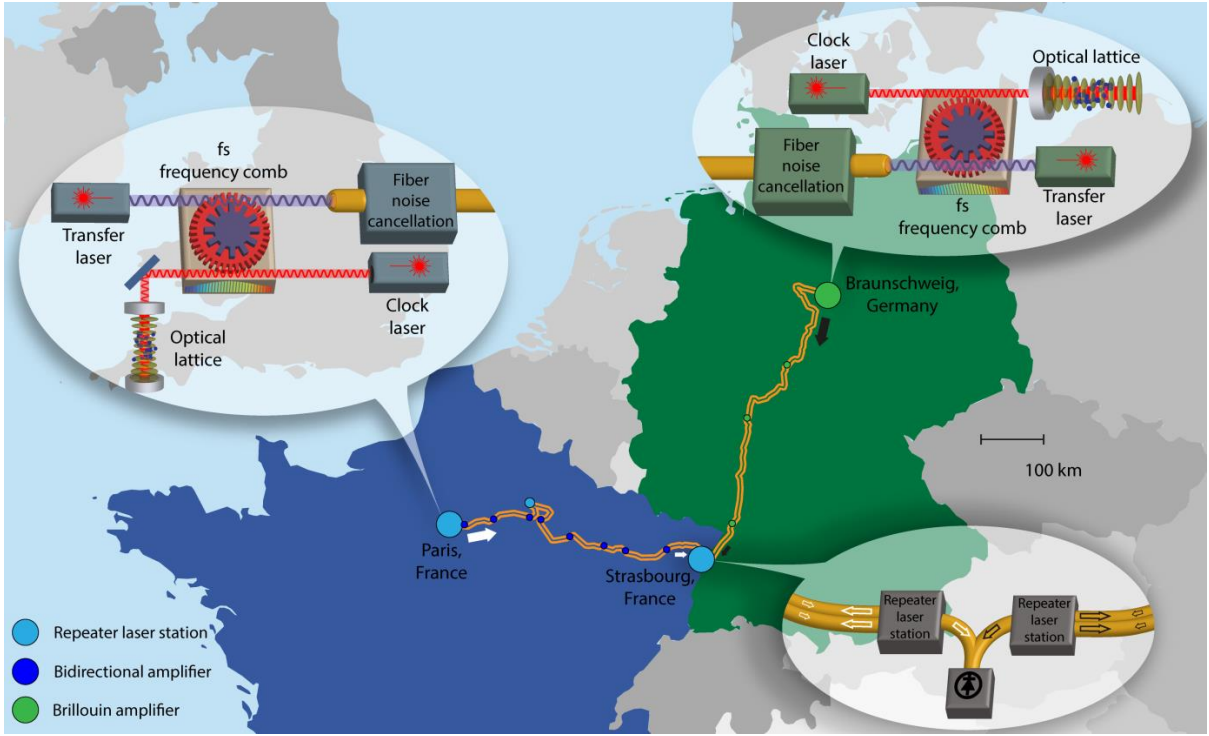

Fig. 1: Schematic of the strontium lattice clock comparison between the national metrology institutes SYRTE and PTB in Paris and Braunschweig, respectively. The course and lay-out of the fiber link sections to Strasbourg are indicated on the map. Additionally, the individual setups consisting of a clock laser, optical lattice, fs frequency comb, transfer laser, and stabilized link are shown schematically. In Strasbourg, the frequency difference between the transfer lasers is measured. For details see the main text and supplementary online text.

The Sr lattice clocks provide a reproducible and long-term stable optical frequency at 429 THz (698 nm) by locking the frequency of an ultra-stable laser (*24,7*) with sub-Hertz linewidth to the ultra-narrow atomic $^1S_0 - {}^3P_0$ transition of laser-cooled $^{87}$Sr atoms. The atoms are trapped in an optical lattice near the magic wavelength (*25*) (see supplementary online text) and are thus practically immune to motional effects such as Doppler shifts. Physical effects leading to offsets of the laser frequency from the unperturbed atomic frequency are quantified in the uncertainty budgets for both optical clocks (Table 1 and supplementary online text). The Sr $^1S_0 - {}^3P_0$ transition frequency is expected to be realized in Paris and Braunschweig with respective fractional uncertainties of $4\times10^{-17}$ and $2\times10^{-17}$ and instabilities of about $1\times10^{-15}/(\tau/s)^{1/2}$ and $5\times10^{-16}/(\tau/s)^{1/2}$ for averaging times $\tau > 10$ s. The details of the actual implementation of state preparation and interrogation are quite different in the two experimental realizations of a Sr lattice clock (*7,19*), which leads to highly uncorrelated results.

Table 1: Uncertainty budget. The numbers vary slightly over the course of the measurement.

| Clock uncertainty | Sr lattice clock Paris | | Sr lattice clock Braunschweig | |
|---|---|---|---|---|
| Effect | Corr. ($10^{-17}$) | Unc. ($10^{-17}$) | Corr. ($10^{-17}$) | Unc. ($10^{-17}$) |
| First and higher order lattice LS | 0 | 2.5 | −1.1 | 1.0 |
| Black-body radiation | 515.5 | 1.8 | 492.9 | 1.3 |
| Black-body radiation oven | 0 | 1.0 | 0.9 | 0.9 |
| Density shift | 0 | 0.8 | 0 | 0.1 |
| Quadratic Zeeman shift | 134.8 | 1.2 | 3.6 | 0.15 |
| Line pulling | 0 | 2.0 | 0 | ≪ 0.1 |
| **Total clocks** | **650.3** | **4.1** | **496.3** | **1.9** |

| Ratio Sr$_{PTB}$/Sr$_{SYRTE}$ | Campaign I Unc. ($10^{-17}$) | Campaign II Unc. ($10^{-17}$) |
|---|---|---|
| Systematics Sr$_{SYRTE}$ | 4.1 | 4.1 |
| Systematics Sr$_{PTB}$ | 2.1 | 1.9 |
| Statistical uncertainty | 2 | 2 |
| fs combs | 0.1 | 0.1 |
| Link uncertainty | < 0.1 | 0.03 |
| Counter synchronization* | 10 | < 0.01 |
| Gravity potential correction** | 0.4 | 0.4 |
| **Total clock comparison** | **11.2** | **5.0** |

LS: light shift; Corr.: fractional correction; Unc.: fractional uncertainty

* Frequency counters have been synchronized in the second campaign.
** The applied gravity potential correction is $-247.2 \times 10^{-17}$.

Measuring the frequency ratio of the two lattice clocks involves two parts: optical frequency combs at each institute compare the local clock laser's frequency with a narrow linewidth transfer laser in the telecommunication band at 194.4 THz (1542 nm) (26). By counting the rf (radio frequency) beat-note signals of the lasers with the frequency combs, the optical frequency ratios between each Sr clock and the associated transfer laser are determined accurately with less than $10^{-18}$ uncertainty (7,19). The transfer lasers are injected from Paris and Braunschweig into two long-haul coherent fiber links, both ending at the University of Strasbourg. There, the beat note between the two transfer lasers is recorded, thus enabling an uninterrupted frequency transfer between the institutes. The beat measurement is performed with an uncertainty of $2\times10^{-20}$.

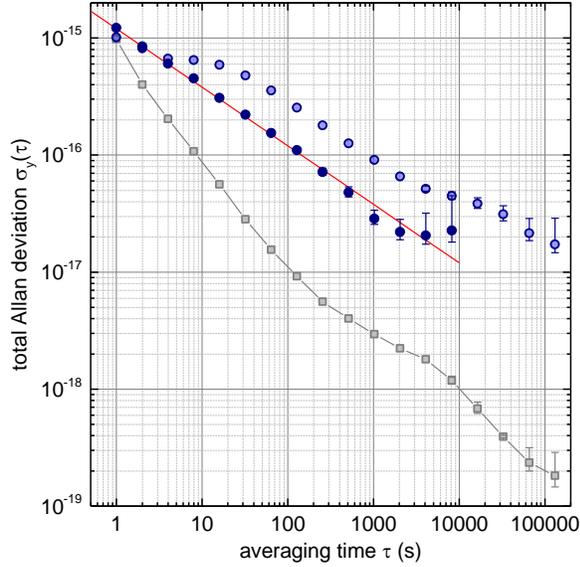

Fig. 2: Instability of the clock comparison expressed as total Allan deviation $\sigma_y$, as a function of averaging time $\tau$ for the first (full circles; MJD 57092) and second measurement campaign (open circles; MJD 57188 – 57198). We attribute the slightly different instabilities to different clock performances. An instability of $1.2\times10^{-15}/(\tau/s)^{1/2}$ is observed during the first campaign (red line). A statistical measurement uncertainty of $2\times10^{-17}$ is thus achieved after only 3000 s measurement time. The squares with connecting lines show the link's contribution to the statistical uncertainty for the data recorded in the 2$^{nd}$ campaign (see supplementary text).

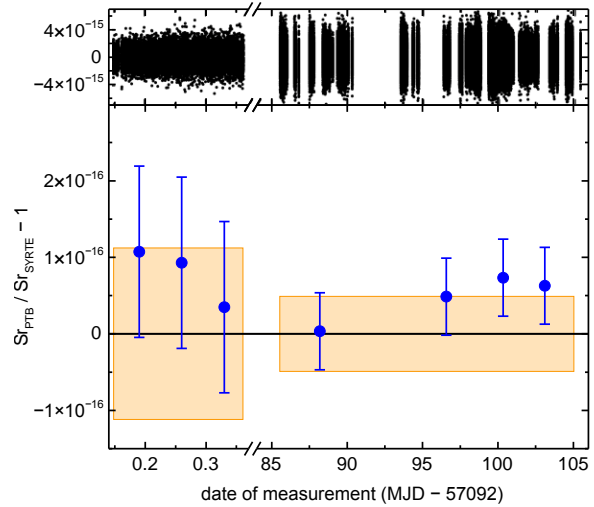

Fig. 3: Frequency ratio ($Sr_{PTB} / Sr_{SYRTE} - 1$). Top: time trace; bottom: aggregate segments of 6000 s length (1$^{st}$ campaign) and 125000 s (2$^{nd}$ campaign). Error bars include statistical and systematic uncertainties (Table 1). Note that the systematic uncertainty is dominating and is not reduced by averaging. On average, the clocks agree within the $\pm1\sigma$ interval of uncertainty (shaded area) around the expected result of zero demonstrating the very good agreement between the two systems.

The main challenge is to ensure that the coherent phase of the transfer laser is preserved over the total distance of the link, despite the phase noise imprinted on the light by environmentally induced optical path length fluctuations of the fiber (*8,10,22,23*). For this purpose, part of the light reaching the remote end of the link is sent back through the same fiber; thus the roundtrip phase noise is detected at the sender's position and actively cancelled, and we accurately transfer the frequency at the input of the fiber to the remote end (supplementary online text).

The power in each optical fiber path is attenuated by about 20 orders of magnitude (205 dB in France, 178 dB in Germany); the attenuation is mostly compensated by bidirectional amplifiers (supplementary online text). The experiment combines two different approaches: in France, repeater laser stations (RLSs) and broad-band amplifiers transfer an optical frequency in parallel with internet traffic (*22*), while in Germany we use a small number of narrow-band, high-gain amplifiers (*23*) on a dedicated fiber. Both links are established as a cascaded up- and down-link in a loop-back configuration to and from Strasbourg, to allow for assessment of the accuracy of the frequency transfer. This is achieved using RLSs (*9,22*) in Strasbourg, which also compare the signals from the two links at Strasbourg (see Fig. 1).

The relative frequency instabilities of the French and German links are $8\times10^{-16}$ and $1\times10^{-15}$, respectively, at one second integration time. They average down faster than the clocks (Fig. 2) and thus contribute negligibly. The fractional uncertainty of the whole link is assessed to be $2.5\times10^{-19}$ (*22,23*). Such long distance links with uncertainties of the order of $10^{-19}$ open the route in the future for comparisons of even more accurate clocks and novel high precision experiments.

The two Sr lattice clocks are separated by 690 km in line-of-sight and by 1415 km of actual fiber links. Their frequency ratio $Sr_{PTB}$ / $Sr_{SYRTE}$ is determined accurately by combining the frequency ratio measurements from Paris, Strasbourg and Braunschweig. The combination of such measurements requires an accurate synchronization of the counters to avoid an additional frequency offset and instability contribution (*8,27*) (see supplementary online text). In this setup we performed two sessions of measurements in March and June 2015. We present here 145 hours of accumulated data.

The instability of the fractional frequency ratio observed between $Sr_{SYRTE}$ and $Sr_{PTB}$ is represented by the total Allan deviation in Fig. 2. The frequency difference is corrected for offsets due to systematic frequency shifts and biases due to synchronization offsets of the counters (Table 1 and supplementary online text). At the $10^{-17}$ level and beyond, besides the global long-wavelength and eventually temporal gravity field variations, especially the local spatial influence of the Earth's gravity potential on the clock frequency is a key aspect for remote clock comparisons. Hence a state-of-the-art determination of the gravity potential at both clock sites becomes necessary to correct for the differential gravitational redshift. For our clock comparisons, the applied average correction was $(-247.4 \pm 0.4)\times10^{-17}$.

A long-term averaging behavior consistent with a white frequency noise level of about $1\times10^{-15}/(\tau/s)^{1/2}$ for the first run and $3\times10^{-15}/(\tau/s)^{1/2}$ for the second one is observed. This confirms that the link instability is negligible for these times. After less than an hour of averaging, we reach a statistical uncertainty of $2\times10^{-17}$. We emphasize that this level of precision is unprecedented for remote clock measurements and not achievable by any other present-day means of remote frequency comparison.

Already this first international all-optical clock measurement surpasses the accuracy of the best possible comparison achievable by measurements against primary cesium clocks by one order of magnitude (*7,19*). In Fig. 3, we show the averaged deviation of the frequency ratio $Sr_{PTB}$

/ Sr$_{SYRTE}$ from unity for segments of the measurement for both campaigns. Their respective lengths of 6000 s and 125000 s are chosen to be close to the maximum averaging time for which we can determine the statistical measurement uncertainty (Fig. 2). For both measurement sessions we find agreement between the two fully independent and remote optical clocks within the combined uncertainties. For the more extensive second campaign, the fractional offset between the two clocks is $(4.7\pm5.0)\times10^{-17}$. Showing optical clock agreement over long distances is an important step towards a redefinition of the SI-second. In addition, the measurement time required for a comparable accuracy is reduced by more than two orders of magnitude compared to measurements with primary clocks. This allows investigating time-dependent (e.g. diurnally varying) clock data, opening up completely new research capabilities (*7,12,14*).

Beyond the results presented here, and taking advantage of an ongoing refinement of fiber links and optical clocks, the addition of further institutes will establish a continental, clock science fiber network. Even for greatly improved optical clocks, running continuously with an instability below $10^{-16}/(\tau/s)^{1/2}$, our fiber link instability still does not pose a limitation due to its faster averaging behavior. Such long distance clock comparison will soon have reached sufficiently high accuracy that the uncertainty of the geo-potential of mid-$10^{-18}$ will be within the measurement range of optical frequency standards. This will turn them into precise and accurate height monitors at the centimeter level over continental distances and paves the path towards a new height reference frame based on clocks.

**Acknowledgments:** We would like to thank T. Bono and L. Gydé for their support and for facilitating the access to the RENATER network and facilities, P. Gris and B. Moya for helping us to establish the cross-border link between Kehl and Strasbourg and for hosting the experiment at the IT center of Uni. Strasbourg, as well as O. Bier from ARTE for support at Kehl and C. Grimm from Deutsches Forschungsnetz (DFN) and W.-Ch. König (Gasline GmbH) for support at the German link. We thank L. Timmen and C. Voigt from the Institut für Erdmessung, Leibniz Universität Hannover, for supporting the geo-potential determination. We acknowledge funding support from the Agence Nationale de la Recherche (ANR blanc LIOM 2011-BS04-009-01, Labex First-TF ANR-10-LABX-48-01, Equipex REFIMEVE+ ANR-11-EQPX-0039), the European Metrology Research Programme (contract SIB-02 NEAT-FT and SIB-55 ITOC), Centre National d'Études Spatiales (CNES), Conseil Régional Île-de-France (DIM Nano'K), CNRS with Action Spécifique Gravitation, Références, Astronomie, Métrologie (GRAM) and the German Research Foundation DFG within RTG 1729 and CRC 1128 geo-Q. The EMRP is jointly funded by the EMRP participating countries within EURAMET and the European Union. Research at PTB was supported by the Centre of Quantum Engineering and Space-Time Research (QUEST).

**Author contributions** G.G. and P.E.P. prepared and coordinated the measurement campaigns with contributions from C.L.; N.Q, N.C, F.S, F.W., G.S., A.A.K., P.E.P. and O.L. built the repeater laser stations, designed, operated, and installed the French fiber link; S.R., G.G., and H.S. designed the German fiber link; S.R., A.Ko., A.Ku., S.K., and G.G. installed, operated, and evaluated the German fiber link; O.L., N.Q., C.C., F.M., S.R., A.Ku., A.Ko., G.G., E.C., and P.E.P designed and implemented the interconnection in Strasbourg; J.L.R., S.B., C.S., E.B., J.L., and R.L.T. evaluated and operated $Sr_{SYRTE}$; S.D., A.A.M., and C.L. evaluated and operated $Sr_{PTB}$; D.N., M.A., M.L., Y.L.C., R.L.T., G.S. built the SYRTE transfer laser and de-drift system; T.L. and U.S. built the PTB transfer laser and de-drift system with contributions from S.R., G.G., and A.Ko.; S.H. built the PTB clock laser; D.N., R.L.T., and Y.L.C. operated the fs combs at LNE-SYRTE; C.G. the fs comb at PTB; H.D determined the geo-potentials at the clock sites; C.G., J.L., R.L.T., G.G., and C.L. analyzed the data. C.L., C.G., J.L., S.R., G.G., D.N., F.S., and P.E.P. resolved counter synchronization issues; S.R. measured the counter offsets between PTB and Strasbourg; C.C. initiated the partnership with RENATER; C.L. wrote the paper with support from P.E.P., A.A.K., G.S., S.R., J.L., R.L.T, and G.G. All authors discussed the results and commented on the paper.

**Supplementary Materials:**

Methods:

*Optical lattice clock operation*

*Optical lattice clock uncertainty evaluation*

*Frequency counter data recording*

*Correction for gravity potential difference*

*Link noise cancellation*

*Counter synchronization and correlation analysis*

*Instability and uncertainty contribution by the link*

*References (28 – 30)*

# Supplementary Materials:

## Methods

*Optical lattice clock operation:*

Both lattice clocks interrogate a few hundred atoms trapped inside a vacuum chamber in an optical lattice that is operated at or close to the Stark shift cancellation wavelength at 813 nm of the Sr clock transition $^1S_0 - {}^3P_0$ at 698 nm or 429 THz (*7,19*). The lattice orientation is vertical in $Sr_{SYRTE}$, while a nearly horizontal lattice is used in $Sr_{PTB}$. To load the atoms into the lattice, Zeeman slowing of a Sr atomic beam and laser cooling in a magneto-optical trap (MOT) operated on the $^1S_0 - {}^1P_1$ transition are performed. The atoms in the MOT are either loaded into a deep lattice, which is compatible with the millikelvin temperature of the atoms and cooled in the lattice on the intercombination line $^1S_0 - {}^3P_1$ ($Sr_{SYRTE}$); or they are first transferred to a second-stage MOT operating on the same transition for further cooling and then loaded in a shallow lattice ($Sr_{PTB}$).

The atoms are alternatingly state-prepared in the stretched Zeeman levels $|m_F| = 9/2$, from which $\Delta m_F = 0$ transitions are driven by Rabi interrogation with a pulse length of 580 ms in a small magnetic bias field ($Sr_{PTB}$) or a pulse length of 200 ms in a larger magnetic bias field of $1.6 \times 10^{-4}$ T ($Sr_{SYRTE}$). Light to drive the 698 nm transition is provided by diode-laser systems that are frequency-stabilized to high-finesse optical resonators (*24*). Each Zeeman transition in sequence is interrogated approximately at both half-width points. Through comparison of the detected average excitation probabilities of the atomic sample on both sides of the line, an error signal is derived to steer the frequency of the interrogation laser to the atomic transition frequency. Interrogation of the $m_F = \pm 9/2$ levels effectively cancels the linear Zeeman shift and measures the magnetic field at the same time. The cooling, preparation and interrogation sequence requires about 1 s.

Light from the interrogation laser is also sent to a fs frequency comb that phase-coherently bridges the frequency gap between the 429 THz clock transition and the transfer lasers at 194.4 THz transmitted through the fiber links.

*Optical lattice clock uncertainty evaluation:*

The residual lattice light shift is evaluated by running the clock successively with different trap depths spanning from 50 $E_R$ to 500 $E_R$ (SYRTE) or 70 $E_R$ to 150 $E_R$ (PTB), where $E_R$ is the lattice recoil energy. Higher-order light shifts (*28*) are accounted for by a non-linear regression (SYRTE) or a correction of the data (PTB). The density shift is obtained by interleaving high density (about 5 atoms per site) and low density (about 2 atoms per site) sequences. No frequency shift is observed between these sequences within the statistical resolution. The quadratic Zeeman shift is evaluated by fitting the clock frequency as a function of the magnetic field obtained from the splitting between the two Zeeman components. The probe light shift is evaluated by a theoretical calculation of the AC polarizability of the Sr clock levels. The AOM phase chirp is both measured by interferometric measurements and by measuring the shift induced in the clock frequency. Line pulling is evaluated by running the clock with different linewidths (SYRTE) or by evaluating possible spurious excitation amplitudes (PTB). To evaluate the black-body radiation shift, calibrated Pt100 temperature sensors are placed around the vacuum chamber spanning the coldest and hottest points. The temperature spreads are 0.9 K (SYRTE) and 0.6 K (PTB), corresponding to uncertainties of 0.26 K and 0.17 K.

*Frequency counter data recording:*

Depending on the dominant noise type, the Allan deviation representing the statistical measurement uncertainty follows characteristic power laws for the temporal averaging of the recorded signals. High suppression of phase noise as present in the fiber link can be achieved by appropriate weighting functions for averaging the frequency data recorded by the dead-time-free counters (*29*). Therefore, we operate all counters in a Λ-averaging mode with 1 s gate time, which effectively represents a low-pass filter. The Allan deviation in Fig. 2 shows an averaging behavior which is consistent with white frequency noise. Thus, a uniformly (or Π-) weighted averaging of the 1 s-Λ-averaged data leads to a frequency average with a smaller statistical uncertainty that is represented by the Allan deviation at the given averaging time τ. (*29*)

*Correction for gravity potential difference:*

Clocks experience a fractional frequency shift that is proportional to $W/c^2$ with the gravity potential $W$ (including a gravitational and a centrifugal component) and the speed of light c. The gravity potential difference between the position of the atoms in each lattice clock and a nearby reference marker was found from spirit leveling and the measured local gravity acceleration of the Earth, while the gravity potential of the reference markers was determined by GPS measurements of the height with respect to a rotational ellipsoid in combination with a geoid model refined by local gravity measurements. In this way we are able to correct the redshift between both clocks with an uncertainty equivalent to about 4 cm in height, considering the accuracy of the geoid model as well as the GPS and leveling measurements. The differential temporal variation of the gravity potential by tides will become relevant in the mid-$10^{-18}$ range of uncertainty for clocks separated as in the experiment here. The absolute variation of the redshift can, however, produce fractional frequency changes relative to an ideal reference of few $10^{-17}$.

*Link noise cancellation:*

Phase-coherent frequency transmission through the fiber link is achieved by active cancelation of the propagation noise originating from environmentally induced fluctuations of the fiber's index of refraction (*8,9*). The round-trip fiber noise is measured at one end. When the light goes forth and back through the same fiber, as in our setup, the corresponding fiber noise contributions are equal and can be rejected very well, even for long-haul fiber links (*8*,*22*,*23*). We achieve high signal-to-noise detection by using heterodyne detection techniques that allows us to remove the influence of parasitic reflections and scattering.

The 705 km-long optical link from SYRTE to Strasbourg uses fibers of the French academic network with parallel data traffic. RLSs (*9*,*22*) repeat the optical phase of an ultra-stable laser operated at 194.4 THz (1542 nm) and build a cascaded link of two spans from Paris to Reims and Reims to Strasbourg. Each RLS contains a laser that is phase-locked to the incoming signal and sent both to the previous and the next link for phase-noise compensation. To ensure continuous and bidirectional propagation along the fibers, the telecommunication nodes and the amplifiers are bypassed. In parallel, a similar down-link from Strasbourg to Paris is established to check the integrity of the frequency dissemination. The 710 km-long optical link from Braunschweig to Strasbourg uses fiber Brillouin amplification (*23*) and a RLS at Strasbourg to cascade this up-link with a parallel down-link from Strasbourg to Braunschweig for evaluation purposes.

A beat note between the RLSs at Strasbourg serving on the links to Paris and Braunschweig is generated and the difference frequency of the two transfer lasers is counted with respect to a low-noise ultra-stable RF oscillator disciplined to a GPS signal. The GPS receiver also delivers a pulse-per-second (PPS) signal by which the counters in Strasbourg are synchronized. The counters in Paris and Braunschweig are synchronized to PPS signals derived from UTC(OP) and UTC(PTB).

*Counter synchronization and correlation analysis:*

During the first measurement campaign, the gate intervals of the remotely operated frequency counters in Strasbourg and Paris were accidentally not synchronized with local realizations of the Coordinated Universal Time (UTC). Via the fiber link, only the synchronization between the counters in Braunschweig and Strasbourg could be measured accurately (*27*). Therefore, the synchronization between the counters in Paris and in Strasbourg was derived indirectly.

In case of a frequency drift $\dot{\upsilon}$ of a transfer laser and a time offset of the counter gates $\Delta T$, a frequency offset $\dot{\upsilon} \cdot \Delta T$ will be found. The infrared lasers used for frequency transfer exhibited a drift of up to 2 Hz/s. Furthermore, the lasers' noise will not be fully canceled from the measurement since the measurements are not fully correlated. This can be used to retrieve $\Delta T$ from the data (*30*) by determining the amount of noise as represented by the Allan deviation $\sigma_y(\tau = 1\text{ s}, 2\text{ s})$ for counter data sets that are interpolated on a dense temporal grid and are shifted with respect to each other. The Allan deviations $\sigma_y$ show a pronounced minimum for the actual $\Delta T$ as was tested for the data measured in Braunschweig and Strasbourg. In this way we could retrieve the missing counter synchronization between Paris and Strasbourg with an uncertainty of 10 ms. The data analysis was then performed based on the shifted datasets such that no further correction had to be applied. Together with the drift rate ủ of the transfer laser between Paris and Strasbourg, this uncertainty causes the counter synchronization uncertainty of $10^{-16}$ in Table 1 for the first measurement.

*Instability and uncertainty contribution by the link:*

To characterize the link uncertainty contribution to the clock comparison, the links were operated using an up- and down-link, in a loop configuration. The proper operation of the link is validated analyzing the frequency offset between the light injected to the up-link with respect to the light received back from Strasbourg on the down-link. This gives an upper limit for the instability and uncertainty of the frequency delivered to Strasbourg.

German link data was selected for the valid time intervals of the clock comparison and thus contains interruptions (Fig. 3). For this data set, we calculated the total Allan deviation: this represents the statistical measurement uncertainty contribution of this part of the link. For the French link, we deduce a slightly smaller statistical uncertainty from shorter but similar recordings, and from the detailed discussion in reference 22. The short connection between the two RLSs in Strasbourg contributes negligibly to the overall link uncertainty. Therefore, the total statistical uncertainty is well represented by the data obtained on the German section, and its total Allan deviation. This data is shown in Fig. 2.

The statistical uncertainty of the zero-compatible frequency offset measured for the German link is given by data in Fig. 2. Thus the systematic uncertainty of the German link during the second campaign is $< 1.5 \times 10^{-19}$. The French link has been characterized to show a systematic uncertainty $< 2 \times 10^{-19}$. The overall link systematic uncertainty is calculated as the quadratic sum of the German and French part, yielding a value of $< 2.5 \times 10^{-19}$.